\documentclass[a4paper,11pt]{article}
\usepackage[nointlimits]{amsmath}
\usepackage{amsfonts}
\usepackage{amssymb}
\usepackage{amsmath}
\usepackage{amsthm}
\usepackage{latexsym}
\usepackage{graphicx}
\usepackage{color}
\usepackage{layout}
\usepackage[english]{babel}
\numberwithin{equation}{section} 

\newcommand{\chr}[2]{\displaystyle\genfrac{\{}{\}}{0pt}{}{#1}{#2}}
\def\AA{{\cal A}}
\def\BB{{\cal B}}
\def\CC{{\cal C}}
\def\costante{C}
\def\DD{{\cal D}}
\def\em{\epsilon_m}
\def\ea{\epsilon_J}
\def\FF{{\cal F}}
\def\GG{{\cal G}}
\def\HH{{\cal H}}
\def\incr{{\delta}}
\def\massasole{{M}}
\def\omt{\widetilde{\omega}}
\def\primoindice{\lambda}
\def\pertx{{F_x}}
\def\perty{{F_y}}
\def\pertz{{F_z}}

\def\secondoindice{\mu}
\def\terzoindice{\nu}
\def\torparam{t}
\newcommand{\nada}[1]{}

\begin{document}

\title{\bf Constraining spacetime torsion with LAGEOS}

\author{
Riccardo March
\\
{\small Istituto per le Applicazioni del Calcolo, CNR,
Via dei Taurini 19, 00185 Roma, Italy},
\\
{\small and INFN - Laboratori Nazionali di Frascati (LNF), via E. Fermi 40, Frascati 00044 Roma, Italy.}
\\
{\small E-mail: r.march@iac.cnr.it}
\\
\\
Giovanni Bellettini
\\
{\small Dipartimento di Matematica, Universit\`a
di Roma ``Tor Vergata'',}
\\
{\small via della Ricerca Scientifica 1, 00133 Roma, Italy},
\\
{\small and INFN - Laboratori Nazionali di Frascati (LNF), via E. Fermi 40, Frascati 00044 Roma, Italy.}
\\
{\small  E-mail:
Giovanni.Bellettini@lnf.infn.it}
\\
\\
Roberto Tauraso
\\
{\small Dipartimento di Matematica, Universit\`a
di Roma ``Tor Vergata'',}
\\
{\small  via della Ricerca Scientifica 1, 00133 Roma, Italy},
\\
{\small and INFN - Laboratori Nazionali di Frascati (LNF), via E. Fermi 40, Frascati 00044 Roma, Italy.}
\\
{\small E-mail: tauraso@mat.uniroma2.it}
\\
\\
Simone Dell'Agnello
\\
{\small INFN - Laboratori Nazionali di Frascati (LNF),
via E. Fermi 40, Frascati 00044 Roma, Italy.}
\\
{\small  E-mail: Simone.Dellagnello@lnf.infn.it}
}
\date{}
\maketitle
\thanks{}
\begin{abstract}
We compute the corrections to the orbital Lense-Thirring effect
(or frame-dragging) in the
presence of spacetime torsion.
We analyze the motion of
a test body in the gravitational field of a rotating axisymmetric
massive body,
using the parametrized framework of Mao, Tegmark, Guth
and Cabi.
In the cases of autoparallel and extremal trajectories,
we derive the specific approximate expression
of the corresponding system of ordinary
differential equations, which are then solved
with methods of Celestial Mechanics.
We calculate the secular variations of the longitudes of the
node and of the pericenter.
We also show how the LAser GEOdynamics Satellites (LAGEOS) can be used to
constrain torsion parameters.
We report the
experimental constraints obtained using both the nodes
and  perigee measurements of the orbital Lense-Thirring effect.
This makes LAGEOS and Gravity Probe B (GPB) complementary frame-dragging and
torsion experiments, since they constrain three different combinations of
torsion parameters.

\end{abstract}


\section{Introduction}\label{sec:introd}
In recent years a lot of effort has been devoted to measure gravitomagnetic
effects due to Earth's rotation \cite{CiWh:95}, \cite{Wi:93},
\cite{Wi:06} predicted by the
theory of General Relativity (GR). In particular, the Lense-Thirring effect
on the orbital motion of a test body can be measured by using the
satellite laser ranging (SLR) technique, whose  data are provided by the
ILRS\footnote{International Laser Ranging Service; see http://ilrs.gsfc.nasa.gov/.}.
By analyzing the laser ranging data
of the orbits of the satellites LAGEOS and LAGEOS II, a measurement
of the Lense-Thirring effect was obtained by Ciufolini and Pavlis
\cite{CiPa:04}.

SLR missions can also be useful to test modifications of GR, such as torsion theories
of gravity.
 A class of theories allowing the presence of torsion is
based on  Riemann-Cartan spacetime, which is
 endowed with
a metric $g_{\secondoindice \terzoindice}$ and a compatible connection.
The resulting
connection $\Gamma^{\primoindice}_{~\secondoindice\terzoindice}$ turns out to be nonsymmetric, and
therefore it originates a non-vanishing torsion tensor. We refer to
\cite{He:76}, \cite{Ha:02} for the details.

In standard torsion theories the source of torsion is considered to be the
intrinsic spin of matter \cite{He:76}, \cite{Ha:02}, \cite{StYa:79},
\cite{YaSt:80}, which
is negligible when averaged over a macroscopic body.
Therefore spacetime torsion would
be observationally negligible in the solar system.
Nevertheless, in \cite{MaTeGuCa:07} Mao, Tegmark, Guth and Cabi (MTGC)
argue that the presence of detectable
torsion in the solar system should be tested experimentally, rather than derived
by means of a specific torsion model.
For this reason, in \cite{MaTeGuCa:07} a theory-independent framework
based on symmetry arguments is developed, and it is determined by
a set $\torparam_1,\torparam_2, w_1,\dots, w_5$ of seven parameters describing torsion
and three further parameters $\mathcal F,\mathcal G,\mathcal H$ describing the metric.
Here, by theory-independent framework, we mean the following:
the metric and the connection are parametrized, around a massive
body, with the help
of symmetry arguments, without reference to a torsion
model based on a specific Lagrangian
(or even on specific field equations).

This parametrized
framework can be used to constrain $\torparam_1,\torparam_2, w_1,\dots, w_5$ from
solar system experiments. In particular, MTGC suggest that
GPB \cite{Ev:11} is an appropriate experiment for this task, and in
\cite{MaTeGuCa:07} they compute precessions of gyroscopes  and  put
constraints on  torsion parameters from
GPB measurements.
In \cite{{HeOb:07}} Hehl and Obukhov argue that measuring torsion requires
intrinsic spin, and
criticize the approach of MTGC, since GPB gyroscopes
do not carry uncompensated elementary particle spin.
Nevertheless, we accept
the general idea that the precise form of the coupling of torsion to matter
should be tested experimentally, and that actual experimental knowledge leaves room
for nonstandard torsion theories which could yield detectable torsion signals in the solar system.
In the present paper we apply the parametrized framework
developed by MTGC for the computation of
satellites orbits around Earth and we put a different set of constraints on torsion parameters from SLR measurements.

MTGC also address the question of whether  there exists a specific
gravitational Lagrangian fitting in the parametrized framework and yielding a torsion signal detectable
by the GPB experiment. As an example they quote the theory of
Hayashi and Shirafuji (HS) in
\cite{HaSh:79}
 where a massive body
generates a torsion field, and they
propose what they call the Einstein-Hayashi-Shirafuji (EHS)
Lagrangian, interpolating GR and HS Lagrangians in a linear way. However, MTGC consider
only a gravitational Lagrangian in vacuum, so that they cannot derive the
equations of motion of test bodies from the gravitational field equations,
which would require a suitable matter coupling.

The EHS model has been criticized by various authors.
In the paper
\cite{FlRo:07},
Flanagan and Rosenthal show that the
linearized EHS theory becomes consistent only if the coefficients
in the Lagrangian are constrained in such a way that
the resulting predictions coincide with those of GR.
In the paper \cite{PuOb:08}, Puetzfeld and Obukhov derive the equations of motion in the framework
of metric-affine gravity theories, which includes the HS theory, and show
that only test bodies with microstructure (such as spin) can couple to torsion.
In conclusion, the EHS theory does not yield a torsion signal detectable for GPB.
For these reasons, in \cite{MaTeGuCa:07}
the EHS Lagrangian is proposed
not as a viable physical model, but as a pedagogical toy model fitting in the parametrized framework,
and giving an illustration of the constraints that can be imposed on torsion by the GPB experiment.
In the present paper we will not consider such a toy model.

As also remarked by Flanagan and Rosenthal in \cite{FlRo:07}, the failure
of constructing the specific EHS Lagrangian
does not rule out the possibility that there may exist other torsion
theories
which could be usefully constrained by solar system experiments.
Such torsion models should fit
in the above mentioned theory-independent framework,
similarly to a parametrized post-Newtonian framework including
torsion. We remark that the parametrized formalism of MTGC does not take into account the intrinsic spin of matter
as a possible source of torsion, and in this sense it cannot be a general torsion framework.
However, it is adequate for the
description of torsion around macroscopic massive bodies
in the solar system, like planets, being the
 intrinsic spin  negligible when averaged over such bodies.

Therefore we think it is worthwhile to continue the investigation of
observable effects in the solar system of nonstandard torsion models within
the MTGC parametrized formalism, under suitable working assumptions.
In particular, our aim is
to extend the GPB gyroscopes computations made in \cite{MaTeGuCa:07}
to the case of  motion of satellites.

In the present paper we compute the corrections to the orbital
Lense-Thirring effect due to the presence of
spacetime torsion described by $\torparam_1,\torparam_2, w_1,\dots, w_5$.
We consider the
motion of a test body in the gravitational field of a rotating
axisymmetric massive body,
under the assumption of slow motion of the test body.
Since we use a parametrized framework without specifying the coupling of torsion to matter,
we cannot derive the equations of motion of test bodies from the gravitational
field equations. Therefore, in order to compute effects of torsion
on the orbits of satellites, we will work out the implications
of the assumption that the trajectory of a test body is either an
extremal or an autoparallel curve. Such trajectories do not need to coincide when
torsion is present.

As in the original paper of Lense and Thirring \cite{LeTh:18},
we characterize the
motion using the six orbital elements of the osculating ellipse.
In terms of these orbital elements, the equations of motion then
reduce to the Lagrangian planetary equations. We calculate the secular
variations of the longitude $\Omega$ of the node and of the longitude
$\widetilde \omega$ of the pericenter. The computed secular variations
show how the corrections to the orbital Lense-Thirring effect depend on
the torsion parameters, and it turns out that the dependence
is only through $w_1, \dots, w_5$.
The data from the LAGEOS satellites are then used to
constrain the relevant linear combinations of the torsion parameters.
More precisely, we constrain two different linear
combinations of $w_1,\dots,w_5$ by using first the measurements
of the nodes
of LAGEOS and LAGEOS II, and then
the measurements
of the nodes
of LAGEOS and LAGEOS II and of the perigee of LAGEOS II.
In particular,
torsion parameters cannot be constrained by satellite experiments
in the case of
extremal trajectories.

While the torsion perturbations to the Lense-Thirring effect
depend only on  $w_1,\dots, w_5$, it turns out that
another relevant relativistic effect, namely the geodetic precession
(or de Sitter effect),
depends on the parameters $t_1$ and $t_2$, and on a further
parameter $t_3$. This latter parameter is involved in a higher order
parametrization of torsion, which is  necessary for the description of the
geodetic precession effect, while it is not necessary at the order of accuracy
required in the present paper. All computations of orbital geodetic precession with
torsion of a satellite are performed in the companion paper   \cite{MaBeTaDe:10}, to which
we will sometimes refer for details.

The paper is organized as follows.
In Section \ref{sec:spator} we briefly recall the notion of spacetime with
torsion.
In Section \ref{sec:geodet} we discuss the case of extremal trajectories.
In Section \ref{sec:autopa} we analyze the equations of autoparallel
trajectories and derive the related system of ordinary differential
equations to first order. The expression of the system clearly reveals
the perturbation due to torsion with respect to the Lense-Thirring
equations.
In Section \ref{sec:comput} we derive the time evolution of
the orbital elements, by applying the classical perturbation theory
of Celestial Mechanics, in particular the Gauss form of the Lagrange planetary
equations.
In Section \ref{sec:correz} we calculate the secular variations
of the orbital elements.
In Section \ref{sec:desitter} we recall some results
from \cite{MaBeTaDe:10}
where torsion solar perturbations are computed. These results will
be used in
 Section \ref{sec:simone}, where we give the observational constraints
that the LAGEOS experiment can place on torsion parameters.
Conclusions are drawn in Section \ref{sec:conclus}.
For convenience of the reader,
in the appendix (Section \ref{sec:append})
we recall from \cite{MaTeGuCa:07} how to
parametrize the metric and torsion tensors, and hence how
to parametrize the connection, under suitable symmetry assumptions.
\section{Spacetime with torsion}\label{sec:spator}
A manifold equipped with a Lorentzian metric
$g_{\secondoindice\terzoindice}$ and a  connection
$\Gamma^\primoindice_{~\secondoindice\terzoindice}$
compatible with the metric
is called a Riemann-Cartan spacetime \cite{He:76}, \cite{Ha:02}. Compatibility means
that  $\nabla_\secondoindice g_{\terzoindice\primoindice}=0$, where $\nabla$ denotes the
covariant derivative. We recall in particular that
for any vector field $v^\primoindice$
$$
\nabla_\secondoindice v^\primoindice \equiv
\partial_\secondoindice v^\primoindice +
\Gamma^\primoindice_{~\secondoindice\terzoindice} v^\terzoindice.
$$
 The connection is determined
uniquely by $g_{\secondoindice\terzoindice}$ and by the torsion tensor
$$
S_{\secondoindice\terzoindice}^{~~~\primoindice} \equiv \frac{1}{2}\left(
\Gamma_{~\secondoindice\terzoindice}^{\primoindice}  -
\Gamma_{~\terzoindice\secondoindice}^{\primoindice}
\right)
$$
as follows:
\begin{equation}\label{full}
\Gamma^{\primoindice}_{\
\secondoindice\terzoindice}=
\chr{\primoindice}{\secondoindice\terzoindice}
- K_{\secondoindice\terzoindice}^{\ \ \ \primoindice},
\end{equation}
where
$\{\cdot\}$ is the Levi-Civita connection, defined by
\begin{equation}\label{levi-civita}
\chr{\primoindice}{\secondoindice\terzoindice}=
\frac{1}{2}g^{\primoindice\rho}\left(\partial_\secondoindice g_{\terzoindice\rho}+
\partial_\terzoindice g_{\secondoindice\rho}-\partial_\rho g_{\secondoindice\terzoindice}\right),
\end{equation}
and
\begin{equation}\label{contorsione}
K_{\secondoindice\terzoindice}^{\ \ \ \primoindice}
\equiv-S_{\secondoindice\terzoindice}^{\ \ \ \primoindice}-S_{\
\terzoindice\secondoindice}^{\primoindice}-S_{\ \secondoindice\terzoindice}^{\primoindice}
\end{equation}
is the contortion tensor.
In the particular case when
$\Gamma^{\primoindice}_{\
\secondoindice\terzoindice}$ is symmetric with respect to
$\secondoindice,\terzoindice$ the torsion tensor
vanishes. We will be concerned here with the case
of nonsymmetric connections
$\Gamma^{\primoindice}_{\ \secondoindice\terzoindice}$.
The case of vanishing torsion tensor corresponds to Riemann
spacetime of GR, while the  case of vanishing
Riemann tensor corresponds to the Weitzenb\"ock spacetime \cite{HaSh:79}.

In the present paper we use the natural gravitational units
$c=1$ and $G=1$.
We will assume that Earth can be approximated as a
uniformly rotating spherical object of mass $m$ and angular momentum $J$.
Following \cite{MaTeGuCa:07}, we use spherical coordinates $(r,\theta,\phi)$ for a satellite
moving in the gravitational field of Earth,
 and we introduce the dimensionless parameters
$\em\equiv m/r$ and $\ea \equiv J/(mr)$. Since the radii of the LAGEOS
orbits (about 6000 km altitude) are much larger than Earth's Schwarzschild
radius, it follows that $\em <<1$. Moreover, since Earth is slowly
rotating, we have $\ea <<1$. Therefore, all computations
will be carried out perturbatively to first  order in $\em$ and $\ea$.

Under spherical axisymmetry assumptions, the metric tensor
$g_{\secondoindice\terzoindice}$ and
the torsion tensor  $S^{~~~\rho}_{\secondoindice\terzoindice}$
have been parametrized to first order in  \cite{MaTeGuCa:07}.
Accordingly,
$g_{\secondoindice\terzoindice}$ is parametrized
 by three parameters $\HH$, $\FF$, $\GG$, and
$S^{~~~\rho}_{\secondoindice\terzoindice}$
is parametrized by seven parameters
$\torparam_1, \torparam_2, w_1,\dots,w_5$,
$$
S^{~~~\primoindice}_{\secondoindice\terzoindice}
= S^{~~~\primoindice}_{\secondoindice
\terzoindice}\left(\torparam_1, \torparam_2,
w_1, \dots, w_5, r, \theta, \phi\right).
$$
Therefore $\Gamma_{~\secondoindice\terzoindice}^{\primoindice}$ becomes an
explicit  function  of all metric and torsion parameters.
It turns out that $t_1, t_2$ contribute to geodetic
precession, while $w_1,\dots,w_5$ contribute to
the frame-dragging precession.
In the Appendix
we report
 the explicit expressions of the parametrized metric and
 torsion tensors, and of the connection, that will be needed in
 the sequel of the paper.

\section{Equations of extremal trajectories}\label{sec:geodet}
In GR structureless test bodies move along geodesics. In a Riemann-Cartan spacetime
there are two different classes of curves, autoparallel and extremal curves, respectively,
which reduce to the geodesics of Riemann spacetime when torsion is zero \cite{He:76}.
Autoparallels are curves along which
the velocity vector is transported parallel to itself
by the connection $\Gamma^\lambda_{~\mu\nu}$. Extremals
are curves of extremal length with respect to the metric $g_{\secondoindice\terzoindice}$.
The velocity vector is transported parallel to itself along
extremal curves by the Levi-Civita connection.
In GR the two types of trajectories
coincide while,
in general, they may differ in presence of torsion.
They are identical when the torsion is totally antisymmetric \cite{He:76},
a condition which is not satisfied within our parametrization.

The equations of motion
of bodies in the gravitational field follow from the field equations due to
the Bianchi identities.
The method of Papapetrou \cite{Pa:51} can be used to derive the equations of motion
of a test body with internal structure, such as for instance a small extended object that may have
either rotational angular momentum or net spin.
In standard torsion theories the trajectories of test bodies with internal structure, in general, are
neither autoparallels nor extremals \cite{He:76}, \cite{Ha:02},
\cite{BaFr:10}, while
structureless test bodies, such as spinless test particles, follow extremal trajectories.

The precise form of the equations of motion
of bodies in the gravitational field depends on the way the matter
couples to the metric and the torsion in the Lagrangian (or in the gravitational field equations).
As explained in the Introduction,
we do not
specify a coupling of torsion to matter, hence we
do not specify the field equations. Moreover,
in our computations of orbits of a satellite (considered as a test body), we will neglect its internal structure.
In a theory-independent framework we cannot derive the equations of motion from the gravitational field equations,
hence
we need some working assumptions on the trajectories of
structureless test bodies: we will investigate the consequences of
the assumption that the trajectories are either extremal or autoparallel curves.
Assuming the trajectory to be an extremal is  natural and consistent with standard torsion theories.
However, extremals depend only on the parameters of the metric,
so that new predictions related to torsion cannot arise.
We will quickly report the computations for the sake
of completeness,
since the metric parameters can be immediately
related to the Parametrized Post Newtonian (PPN) parameters (see
\eqref{postniutonianparamiters}),
and the orbital Lense-Thirring effect in the case of extremal trajectories and a PPN metric is known.

The system of equations of extremal trajectories reads as
\begin{equation}\label{extremal-equation}
\frac{d^2 x^{\primoindice}}{d\tau^2} +
\chr{\primoindice}{\secondoindice\terzoindice}
\frac{d
x^{\secondoindice}}{d\tau}\frac{d x^{\terzoindice}}{d\tau}=0,
\end{equation}
where $\tau$ is the proper time. For slow motion of the satellite
we can make the substitution $d\tau \simeq dt$, so that
$$
\frac{d^2 x^\alpha}{dt^2} + \chr{\alpha}{\secondoindice\terzoindice}
\frac{dx^\secondoindice}{dt}
\frac{dx^\terzoindice}{dt} =0,
$$
for $\alpha \in \{1,2,3\}$.
We assume that the velocity of the satellite is small enough
so that we can neglect the quadratic terms in the velocity.
Then, being $x^0 = t$ we have
\begin{equation}\label{extremal}
\frac{d^2 x^\alpha}{dt^2} +
\chr{\alpha}{00} +
2\chr{\alpha}{0\beta}
\frac{dx^\beta}{dt} =0,
\end{equation}
for $\beta \in \{1,2,3\}$.

All perturbations considered here are so small
that can be superposed linearly. Since we are only interested in
the perturbations due to Earth's rotation, as in the original
Lense-Thirring paper \cite{LeTh:18} we are allowed to neglect the
quadratic terms in the velocities which yield an advance of
the perigee of the satellite. The value of the advance of the perigee
for an extremal orbit and a PPN metric can be found in \cite[Chapter 7, formula (7.54)]{Wi:93}.

We use for $x^\alpha$ spherical coordinates $(r,\theta,\phi)$.
The Levi-Civita connection $\{\cdot\}$
can be obtained from the expression of
$\Gamma^\primoindice_{~\secondoindice\terzoindice}$
given in the Appendix
by setting to zero all torsion parameters $t_1,t_2,w_1,\dots,w_5$.
Substituting the resulting expression in \eqref{extremal}
one gets
\begin{equation}\label{sistODEsferiche-extremal}
\left\{\begin{array}{l}
\ddot{r}r-(\HH\slash 2)\em+\GG\dot{\phi}r \sin^2\theta \em\ea=0,
\\
\\
\ddot{\theta}r-2\GG\dot{\phi}\sin\theta \cos\theta \em\ea=0,
\\
\\
\ddot{\phi}r^2 \sin\theta -
\GG\dot{r} \sin\theta \em\ea+2\GG \dot{\theta} r \cos\theta \em\ea=0.
\end{array}
\right.
\end{equation}
The equations of motions \eqref{sistODEsferiche-extremal} depend neither
on the metric parameter $\FF$ nor on the torsion parameters.
System \eqref{sistODEsferiche-extremal} to lowest order becomes
\begin{equation*}
\frac{d \vec v}{dt} = \frac{\HH}{2} \frac{m}{r^2} \hat e_r,
\end{equation*}
where $\hat e_r$ is the unit vector in the radial direction.
Imposing the Newtonian limit yields
$\HH=-2$ as in a PPN metric (see also \cite[formula (23)]{MaTeGuCa:07}).

We now transform \eqref{sistODEsferiche-extremal} in rectangular coordinates
$x=r \sin\theta \cos\phi$,
$y=r \sin\theta \sin\phi$,
$z=r \cos\theta$.
We compute the second derivatives of $x,y,z$ with respect to time
in the approximation of slow motion. Neglecting all terms containing
squares and products of first derivatives with respect to
$(r,\theta,\phi)$, we get
\begin{equation}\label{hessianappr}
\left\{\begin{array}{l}
\ddot x =
\ddot r \sin\theta \cos\phi
+
\ddot \theta r \cos\theta \cos\phi
-
\ddot \phi r \sin\theta \sin\phi,
\\
\\
\ddot y = \ddot r
\sin\theta \sin\phi +
\ddot \theta r \cos\theta \sin\phi
+
\ddot \phi r \sin\theta \cos\phi,
\\
\\
\ddot z =
\ddot r  \cos\theta -
\ddot \theta r \sin\theta.
\end{array}\right.
\end{equation}
Using \eqref{sistODEsferiche-extremal} and \eqref{hessianappr}
we obtain the following system for the equations of motion:
\begin{equation}\label{equazmoto-extremal}
\left\{\begin{array}{l}
\ddot{x}=
\displaystyle
-{\frac {\em }{{r}^{2}}} x
-\GG{\frac{\em\ea}{r^3}} \Big[
\left({x}^{2}+{y}^{2}-2{z}^{2} \right)\dot{y}+
3 yz\dot{z} \Big],
\\
\\
\displaystyle
\ddot{y}=
-{\frac {\em }{{r}^{2}}} y
+\displaystyle
\GG{\frac{\em\ea}{r^3}} \Big[
\left({x}^{2}+{y}^{2}-2{z}^{2} \right)\dot{x}+
3 xz\dot{z} \Big],
\\
\\
\displaystyle
\ddot{z}=
-{\frac {\em }{{r}^{2}}} z+\displaystyle
\GG{\frac{\em\ea}{r^3}}3z\left(y\dot{x}-x\dot{y}\right).
\end{array}\right.
\end{equation}
Note that when $\GG = -2$ system \eqref{equazmoto-extremal} reduces to the
equations of motion found by the Lense-Thirring \cite[formula (15)]{LeTh:18}.
Hence the relativistic perturbation of the Newtonian force is just multiplied by the factor $-\GG\slash 2$
with respect to the original Lense-Thirring equations. It follows that the formulae of
precession of the orbital elements of a satellite can be obtained by multiplying the original
Lense-Thirring formulae \cite[formula (17)]{LeTh:18} by the factor $-\GG\slash 2$.
The details of the computation, based on the
Lagrange planetary equations of Celestial Mechanics, can be also retrieved from the computations for
autoparallel trajectories given in the next sections,
by setting to zero all  torsion parameters
$t_1,t_2,w_1,\dots,w_5$.

Using the standard astronomical notation,
we denote by $\Omega$ the
longitude of the node and by $\omega$ the argument of the
perigee of the satellite's orbit.
The secular contributions to the variations of $\Omega$ and $\omega$ are:
\begin{equation}\label{geodesicsecular}
(\incr \Omega)_{\rm sec} = -
\frac{\GG J}{a^3(1-e^2)^{3/2}}~t,
\qquad
(\incr \omega)_{\rm sec}
=\frac{3\GG J\cos i}{a^3(1-e^2)^{3/2}}~t,
\end{equation}
where $a$ is the semimajor axis of the satellite's orbit, $e$ is the eccentricity, $i$ is the orbital inclination,
and $t$ is time. When $\GG = -2$
the quantities in \eqref{geodesicsecular}
reduce to the classical corresponding Lense-Thirring ones.

Since the expressions of $(\incr \Omega)_{\rm sec}$ and $(\incr \omega)_{\rm sec}$
depend only on $\GG$,
the measurements of satellites experiments cannot be used
to constrain the torsion parameters.

\section{Equations of autoparallel trajectories}\label{sec:autopa}
In standard torsion theories the trajectories of
structureless test bodies follow extremal trajectories \cite{He:76},
\cite{Ha:02},
which depend only on the metric. However,
new predictions related to torsion may arise when considering
the autoparallel trajectories.
In the following we give some motivations
which make worthwhile the investigation of autoparallel trajectories.

Since in spacetime with torsion parallelograms are in general not
closed, but exhibit
a closure failure proportional to the torsion, Kleinert and Pelster argue in
\cite{KlPe:99}
that the variational procedure in the action principle for the motion of structureless test bodies
must be modified.
In the standard variational procedure for finding the extrema of the action,
paths are varied keeping the endpoints fixed in such a way that variations form
closed paths.
However, in the formalism of \cite{KlPe:99},
the closure failure makes the variation at the
final point nonzero, and this gives rise to
a force due to torsion. When this argument is applied to the action principle for structureless test bodies
it turns out that the resulting torsion force changes extremal trajectories to
autoparallel ones
(see \cite{KlPe:99} for the details).
Kleinert and Shabanov find an analogous result in \cite{KlSh:98} where they show that
the geometry of spacetime with torsion can be induced by embedding its curves in a euclidean space
without torsion. Kleinert et al. also argue in
\cite{KlPe:99}, \cite{KlSh:98} that autoparallel
trajectories are consistent with the principle of inertia,
since a structureless test body
will change its direction in a minimal way at each  time, so that the trajectory is as
straight as possible.

The approach of Kleinert et al. has been criticized by Hehl and Obukhov in \cite{{HeOb:07}}
since the equations of autoparallel trajectories have not been derived
from the energy-momentum conservation laws. Kleinert investigates this issue in \cite{Kl:00} and
finds that, due to the closure failure, the energy-momentum tensor
of spinless point particles satisfies a different conservation law
with respect to the one
satisfied in torsion theories such as \cite{He:76}, \cite{Ha:02}.
The resulting conservation law yields autoparallel trajectories for spinless test particles.
Kleinert then addresses the question of
whether this new conservation law allows for the construction of an
extension of Einstein field equations to spacetime with torsion.
The author gives an answer
for the case of torsion derived from a scalar potential (see \cite{Ha:02} for a discussion of
this kind of torsion). In this case the autoparallel trajectories are derived from the gravitational field equations
via the Bianchi identities, though the field equation for the scalar field,
which is the potential of torsion,
is unknown.

In \cite{DeTu:82} Dereli and Tucker show that the theory of Brans-Dicke can be reformulated as a field
theory on a spacetime with dynamic torsion determined by the gradient of the Brans-Dicke scalar field.
Then in \cite{DeTu:02} they
suggest that the autoparallel trajectory of a spinless test particle in
such a torsion geometry
is a possibility that has to be taken into account.
In \cite{DeTu:02}
the autoparallel trajectories of massive spinless test particles are analyzed
in the background of a spherically symmetric, static solution
to the Brans-Dicke theory and the
results are applied to the computations of the orbit of Mercury. In
\cite{CeDeTu:04}
the autoparallel trajectories of spinless particles are analyzed in the background of a
Kerr Brans-Dicke geometry. In \cite{DeTu:04a}, \cite{BuDeTu:04}
the equations of autoparallel trajectories
are derived from the gravitational field
equations and Bianchi identities, in the special case
of matter modeled as a pressureless fluid, and
torsion expressed solely in terms of the gradient of the Brans-Dicke
scalar field.

The above quoted results show that there is an interest
in the autoparallels in spacetime with torsion, which make
worthwhile their investigation in the present paper.
The system of equations of autoparallels reads as
\begin{equation}\label{press}
\frac{d^2 x^{\primoindice}}{d\tau^2} +\Gamma^{\primoindice}_{\
\secondoindice\terzoindice}\frac{d
x^{\secondoindice}}{d\tau}\frac{d x^{\terzoindice}}{d\tau}=0,
\end{equation}
where $\tau$ is the proper time \cite{Po:71}.
Observe that only the symmetric part $\frac{1}{2} (\Gamma_{~\mu\nu}^{\lambda}
 + \Gamma_{~\nu\mu}^{\lambda})$ of the connection enters in \eqref{press};
moreover, starting from \eqref{press} the totally antisymmetric part of
$S_{\lambda\mu\nu}$ cannot be measured.

The trajectory of a test body has to be a time-like curve.
Since the connection is compatible with the metric, the quantity
$g_{\secondoindice\terzoindice}\frac{dx^{\secondoindice}}{d\tau}\frac{d x^{\terzoindice}}{d\tau}$
is conserved by parallel transport. The tangent vector $\frac{dx^{\secondoindice}}{d\tau}$ to the trajectory
undergoes parallel transport by the connection along the autoparallel. Therefore, an autoparallel
that is time-like at one point has this same orientation everywhere, so that the trajectory is strictly contained in
the light cone determined by $g_{\secondoindice\terzoindice}$, in a neighbourhood of every
of its points. Hence the compatibility of the connection with the metric ensures
that autoparallels fulfil a necessary requirement for causality.

For slow motion of the satellite
we can make the substitution $d\tau \simeq dt$, so that
$$
\frac{d^2 x^\alpha}{dt^2} + \Gamma_{~\secondoindice\terzoindice}^\alpha
\frac{dx^\secondoindice}{dt}
\frac{dx^\terzoindice}{dt} =0,
$$
for $\alpha \in \{1,2,3\}$.
Again, we assume that the velocity of the satellite is small enough
so that we can neglect the terms which are quadratic in the velocity.
Then, being $x^0 = t$ we have
\begin{equation}\label{malad}
\frac{d^2 x^\alpha}{dt^2} +
\Gamma_{~00}^\alpha +
(\Gamma_{~\beta 0}^\alpha
+\Gamma_{~0 \beta}^\alpha)
\frac{dx^\beta}{dt} =0,
\end{equation}
for $\beta \in \{1,2,3\}$.

As in the previous section, all the perturbations that we are considering here are so small
that can be superposed linearly.
We are allowed to neglect the
quadratic terms in the velocities which yield an advance of
the perigee of the satellite. Such an advance of the perigee
for an autoparallel orbit in presence of torsion
has been computed
in \cite{MaBeTaDe:10}.

We use for $x^\alpha$ spherical coordinates $(r,\theta,\phi)$.
Substituting in \eqref{malad} the expression of
$\Gamma^\primoindice_{~\secondoindice\terzoindice}$ given in the Appendix
one gets
\begin{equation}\label{sistODEsferiche}
\left\{\begin{array}{l}
\ddot{r}r+\CC\em+\DD\dot{\phi}r \sin^2\theta \em\ea=0,
\\
\\
\ddot{\theta}r-\BB\dot{\phi}\sin\theta \cos\theta \em\ea=0,
\\
\\
\ddot{\phi}r^2 \sin\theta +
\AA\dot{r} \sin\theta \em\ea+\BB \dot{\theta} r \cos\theta \em\ea=0,
\end{array}
\right.
\end{equation}
where
\begin{equation}\label{ABCD}
\left\{\begin{array}{l}
\AA=-\GG+ w_1-w_3,
\\
\\
\BB=2\GG+w_2-w_4,
\\
\\
\displaystyle
\CC= \torparam_1-\frac{\HH}{2},
\\
\\
\DD=\GG- w_1-w_5.
\end{array}\right.
\end{equation}
Note that equations of motions \eqref{sistODEsferiche} do not depend
on the metric parameter $\FF$ and on the torsion parameter $\torparam_2$.
Moreover, the dependence on $w_2$ and $w_4$ appears only through
their difference.

System \eqref{sistODEsferiche} to lowest order becomes
\begin{equation*}
\frac{d \vec v}{dt} = - \CC \frac{m}{r^2} \hat e_r,
\end{equation*}
where $\hat e_r$ is the unit vector in the radial direction.
Imposing the Newtonian limit it follows that
(see also \cite[formula (23)]{MaTeGuCa:07})
\begin{equation}\label{limniuti}
\CC=1.
\end{equation}
Since the Newtonian limit fixes the value of $t_1$,
the equations of autoparallels depend only on the
parameters $w_1,\dots,w_5$ (called frame-dragging torsion
parameters in \cite{MaTeGuCa:07}). Therefore the precession
of satellite's orbital elements will depend only
on such torsion parameters, as it has been found in \cite{MaTeGuCa:07}
for gyroscopes.

Using \eqref{sistODEsferiche} and \eqref{hessianappr}
we obtain the following system for the equations of motion:
\begin{equation}\label{equazmoto}
\left\{\begin{array}{l}
\ddot{x}=
\displaystyle
-{\frac {\em }{{r}^{2}}} x
+{\frac{\em\ea}{r^3}} \Big[(\DD+\AA) xy \dot{x}+
\left(-\DD{x}^{2}+\AA{y}^{2}+\BB{z}^{2} \right)\dot{y}+
\left( \AA-\BB\right) yz\dot{z} \Big],
\\
\\
\displaystyle
\ddot{y}=
-{\frac {\em }{{r}^{2}}} y
+\displaystyle
{\frac{\em\ea}{r^3}} \Big[-(\DD+\AA) xy \dot{y}+
\left(-\AA{x}^{2}+\DD{y}^{2}-\BB{z}^{2} \right)\dot{x}-
\left( \AA-\BB\right) xz\dot{z} \Big],
\\
\\
\displaystyle
\ddot{z}=
-{\frac {\em }{{r}^{2}}} z+\displaystyle
{\frac{\em\ea}{r^3}}(\DD+\BB)z\left(y\dot{x}-x\dot{y}\right).
\end{array}\right.
\end{equation}
Note that in case of no torsion (i.e. $w_i=0$ for any $i=1,\dots,5$) and
when $\GG = -2$ system \eqref{equazmoto} reduces to the
equations of motion found by the Lense-Thirring \cite[formula (15)]{LeTh:18}.
\section{Computation of orbital elements via perturbation
theory}\label{sec:comput}
The system \eqref{equazmoto} expressing the motion along
autoparallel trajectories can be  written in the form
\begin{equation}
\left\{\begin{array}{l}
\ddot{x}=
\displaystyle
 -{\frac {m}{{r}^{3}}}x+\pertx,
\\
\\
\ddot{y}=
\displaystyle
 -{\frac {m}{{r}^{3}}}y+\perty,
\\
\\
\ddot{z}=
\displaystyle
 -{\frac {m}{{r}^{3}}}z+\pertz,
\end{array}\right.
\end{equation}
where $(\pertx, \perty, \pertz)$ is the perturbation with respect to
the Newton force,
\begin{equation}\label{exprpert}
\left\{\begin{array}{l}
\pertx=
\displaystyle
{\frac{ma}{r^5}} \Big[(\DD+\AA) xy \dot{x}+
\left(-\DD{x}^{2}+\AA{y}^{2}+\BB{z}^{2} \right)\dot{y}+
\left( \AA-\BB\right) yz\dot{z} \Big],
\\
\\
\perty=
\displaystyle
{\frac{ma}{r^5}} \Big[-(\DD+\AA) xy \dot{y}+
\left(-\AA{x}^{2}+\DD{y}^{2}-\BB{z}^{2} \right)\dot{x}-
\left( \AA-\BB\right) xz\dot{z} \Big],
\\
\\
\pertz=
\displaystyle
{\frac{ma}{r^5}}(\DD+\BB)z\left(y\dot{x}-x\dot{y}\right).
\end{array}\right.
\end{equation}
We use the standard coordinates transformation
\cite{BrCl:61},
\cite{GeWe:71}
used in Celestial Mechanics
$$
\left\{\begin{array}{l}
x=r\left(\cos u \cos\Omega - \sin u \sin\Omega \cos i\right),
\\
\\
y=r\left(\cos u \sin\Omega+\sin u \cos\Omega \cos i\right),
\\
\\
z=r \sin u \sin i,
\\
\end{array}\right.
$$
where $i$ is the orbital inclination, $\Omega$ is the longitude of the
node, and $u$ is the argument of latitude.
The vector $(\pertx,\perty,\pertz)$ can be decomposed in the standard way along
three mutually orthogonal axes  as
\begin{equation}\label{STW}
\left\{\begin{array}{l}
S= \displaystyle
\frac{x}{r} \pertx + \frac{y}{r} \perty + \frac{z}{r} \pertz,
\\
\\
T=\displaystyle
\frac{\partial (x/r)}{\partial u}
 \pertx
+
\displaystyle
\frac{\partial (y/r)}{\partial u}
\perty
+\displaystyle
\frac{\partial (z/r)}{\partial u}
\pertz,
\\
\\
\sin u ~W= \displaystyle
\frac{\partial (x/r)}{\partial
i}
\pertx
+\displaystyle
\frac{\partial
(y/r)}{\partial i}
\perty
+
\displaystyle
\frac{\partial (z/r)}{\partial i}
\pertz.
\end{array}\right.
\end{equation}
Here $S$ is the component along the instantaneous radius vector,
$T$ is the component perpendicular to the instantaneous radius vector in
the direction of motion, and $W$ is the component normal to the
osculating plane of the orbit (colinear with the angular momentum vector).
Then, substituting \eqref{exprpert} into \eqref{STW} gives
\begin{equation}\label{STWreprise}
\left\{\begin{array}{l}
S=\displaystyle
-\frac{J}{r^2}\DD\cos i ~ \dot{u},
\\
\\
T=\displaystyle
-\frac{J}{r^3}\AA\cos i ~ \dot{r},
\\
\\
W=\displaystyle
\frac{J}{r^3}
\sin i \left(\AA \cos u ~ \dot{r}-\BB \sin u r ~ \dot u\right).
\end{array}
\right.
\end{equation}
Note that in case of no torsion and when $\GG=-2$ formulae
\eqref{STWreprise} reduce to the components found by Lense-Thirring (see
equations (16) in \cite{LeTh:18}).

Let us now recall \cite{BrCl:61},
\cite{GeWe:71} that, using the method of variation of constants,
\begin{equation*}
r=\displaystyle
\frac{a(1-e^2)}{1+e\cos v},
\end{equation*}
where $a$ is the semimajor axis of the satellite's orbit, $e$ is the
eccentricity, $v$ is the true anomaly, and
\begin{equation*}
\dot{r}=
\displaystyle
\frac{r^2e\sin v}{a(1-e^2)}\dot{v}, \qquad
r^2\dot{v}=\displaystyle
na^2 (1-e^2)^{1/2},
\end{equation*}
$n = 2\pi/U$, $U$ the period of revolution. Following the standard
astronomical notation, we let $\omega$ be the argument of the
perigee,
and
 $\widetilde \omega = \Omega + \omega$ be the longitude
of the perigee.

We also recall the following planetary equations of Lagrange in
the Gauss form \cite[Ch. 6, Sec. 6]{GeWe:71}:
\begin{equation}\label{gaussform}
\left\{\begin{array}{l}
\displaystyle
\frac{da}{dt}=\frac{2}{n (1-e^2)^{1/2}}
\left[
Se\sin v +T ~ {a(1-e^2)\over r}
\right],
\\
\\
\displaystyle
\frac{de}{dt}=
\frac{(1-e^2)^{1/2}}{na}
\left[
S\sin v +T\left(e+{r+a\over
a}\cos v \right)
\right],
\\
\\
\displaystyle
\frac{di}{dt}=
\frac{1}{na^2 (1-e^2)^{1/2}} ~ W r \cos u,
\\
\\
\displaystyle
\frac{d\Omega}{dt}=
\frac{1}{na^2 (1-e^2)^{1/2} \sin i} ~ W r \sin u,
\\
\\
\displaystyle
\frac{d\omt}{dt}=
\frac{(1-e^2)^{1/2}}{nae}
\left[
-S\cos v +T\left(1+{r\over
a(1-e^2)}
\right)\sin v
\right]+
2\sin^2 \frac{i}{2} ~{d\Omega\over dt},
\\
\\
\displaystyle
\frac{d L_0}{dt}=- \frac{2}{na^2} ~ S r
+
{e^2\over 1+(1-e^2)^{1/2}}
~ \frac{d\omt}{dt}+2
(1-e^2)^{1/2}
\sin^2 \frac{i}{2} ~ {d\Omega\over dt},
\end{array}
\right.
\end{equation}
where $L_0 = -\tau n + \widetilde \omega$ is the longitude at epoch,
and $\tau$ is the time of periapsis passage.

Using the expressions of $S$, $T$ and $W$ given by \eqref{STWreprise}
and integrating the Lagrange planetary equations
we compute the variations of the orbital
elements.
According to perturbation theory, we regard
the orbital elements as approximately constant
in the computation of such integrals.
 Since $u = v + \widetilde \omega-
\Omega$,
we can make use of the approximation
\begin{equation}\label{udotugualevdot}
\dot{u}\simeq
\dot{v}.
\end{equation}

Inserting \eqref{STWreprise}-\eqref{udotugualevdot} into
\eqref{gaussform} yields
\begin{equation}\label{me?}
\left\{\begin{array}{rcl}
\displaystyle
\frac{da}{dt}&=&
\displaystyle
-{2Je\cos i ~(1+e\cos v )^2\sin v \over
na^2(1-e^2)^{5/2}}\left(\AA\dot{v}+\DD\dot{u}\right),
\\
\\
\displaystyle
\frac{de}{dt}&=&
\displaystyle
-{J\cos i \sin v \over
na^3(1-e^2)^{3/2}}
\Big[
e (e+2\cos v +e\cos^2 v )\AA\dot{v}+(1+e\cos v )^2 \DD\dot{u}
\Big],
\\
\\
\displaystyle
\frac{di}{dt}&=&
\displaystyle
{J\sin i \cos u\over
na^3(1-e^2)^{3/2}}
\Big[
e \sin v \cos u\AA\dot{v}-\sin u (1+e\cos v )
\BB\dot{u}
\Big],
\\
\\
\displaystyle
\frac{d\Omega}{dt}&=&
\displaystyle
{J \sin u \over
na^3(1-e^2)^{3/2}}
\Big[
e \sin v \cos u
\AA\dot{v}-\sin u(1+e\cos v)\BB\dot{u}
\Big],
\\
\\
\displaystyle
\frac{d\omt}{dt}&=&
\displaystyle
{J\cos i\over
na^3 e(1-e^2)^{3/2}}
\Big[
(1+e\cos v)^2\cos v\DD\dot{u}- e \sin^2 v
(2+e\cos v)\AA\dot{v}
\Big]
+
2\sin^2 \frac{i}{2} ~ {d\Omega\over dt},
\\
\\
\displaystyle
\frac{d L_0}{dt}&=&
\displaystyle
{2J\cos i\over na^3(1-e^2)}(1+e\cos v) \DD\dot{u}
\displaystyle
+{e^2\over 1+
(1-e^2)^{1/2}
}~ {d\omt\over
dt}
+2
(1-e^2)^{1/2}
\sin^2 \frac{i}{2} ~{d\Omega\over dt}.
\end{array}\right.
\end{equation}
Recalling \eqref{udotugualevdot}, we now integrate
\eqref{me?} with respect to $v$. Therefore we find for the
variations of the orbital elements:
\begin{eqnarray*}
\incr a&=&{2 Je\cos i \cos v\over
na^2(1-e^2)^{5/2}}
~(\AA+\DD)
\left(1+e\cos v+{1\over 3}e^2\cos^2 v\right),
\\
\\
\\
\incr e&=&{J\cos i \cos v\over
na^3(1-e^2)^{3/2}}\left[(\AA+\DD)\left(1+e\cos v+{1\over
3}e^2\cos^2 v \right)
-\AA(1-e^2)\right],
\\
\\
\\
\incr i&=&{J\sin i\over
12na^3(1-e^2)^{3/2}}
\Bigg[
4(\AA+2\BB)e\cos v\cos^2 u-4(\BB+2\AA)
e\cos v
\\
\\
&&
+2(\BB+2\AA)e\sin v\sin(2u)+3\BB\cos(2u)
\Bigg],
\\
\\
\incr \Omega&=& {J\over 6na^3(1-e^2)^{3/2}}
\Bigg\{ -3\BB v+{3\BB\over
2}\sin(2u)
\\
&& +e\Big[
2(\AA-\BB)\sin v+
(\AA+2\BB)\sin(2u)\cos v-2 (2\AA+\BB) \sin v\cos^2 u
\Big]
\Bigg\},
\end{eqnarray*}
\begin{eqnarray*}
\incr \omt
&=&
{J\over n a^3 e(1-e^2)^{3/2}}
 \cos i
~ \Bigg\{
\sin v
    \left[\DD+(\AA+\DD)e\cos v+{1\over
3}(2\DD-\AA)e^2
    \right.
\\
&& \left.
+{1\over 3}(\AA+\DD)e^2\cos^2 v \right]+
(\DD-\AA)ev\Bigg\}
+2\sin^2 \frac{i}{2}~ \incr \Omega,
\\
\\
\incr L_0&=&{2 J\cos i\over na^3 (1-e^2)}\DD\left(v+e\sin v\right)
+{e^2\over 1+
(1-e^2)^{1/2}
} ~ \incr \omt+2
(1-e^2)^{1/2}
\sin^2
\frac{i}{2} ~\incr
\Omega.
\end{eqnarray*}
We note that the contributions of the components $S$ and $T$ to the derivative
$\frac{da}{dt}$ are proportional to $\DD\dot{u}$ and $\AA\dot{v}$, respectively,
with the same proportionality
constant. Using the approximation $\dot{u}\simeq\dot{v}$
it turns out that in the classical Lense-Thirring case, where the torsion parameters vanish and $-\AA=\DD=\GG$,
there is a cancellation of such contributions
in such a way that $\incr a$ vanishes. Conversely, in presence of torsion, if the eccentricity of the orbit
is nonzero, the contributions of the radial and of the tangential component of the perturbative force
differ,
so that $\incr a$ does not vanish, yielding a periodic perturbation of the semimajor axis
of the satellite's orbit.

\section{Torsion corrections to the Lense-Thirring effect}\label{sec:correz}
We observe that only periodic terms appear in $\incr a$, $\incr e$
and $\incr i$.
Secular terms appear in $\incr \Omega$, $\incr \widetilde \omega$
and $\incr L_0$. Since $v=nt + {\rm periodic~terms~in}~v$, the
secular contributions to the variations of the corresponding
orbital elements are:
\begin{equation}\label{secula}
\left\{
\begin{array}{rcl}
(\incr \Omega)_{\rm sec} &=&
\displaystyle
-{J\over 2a^3(1-e^2)^{3/2}}~ \BB t,
\\\\
(\incr \omt)_{\rm sec}
&=&
\displaystyle
{J\over a^3(1-e^2)^{3/2}}
\left[
\DD-\AA-(\BB+2\DD-2\AA)\sin^2 \frac{i}{2}
\right]
t,
\\\\
(\incr L_0)_{\rm sec} &=&
\displaystyle
\frac{J}{a^3(1-e^2)}
\Bigg\{
2  \DD
+{e^2\over 1+(1-e^2)^{1/2}
}
~
{1\over (1-e^2)^{1/2}}
\left[
\DD-\AA-(\BB+2\DD-2\AA)\sin^2 \frac{i}{2}
\right]
\\\\
&& - (\BB + 4 \DD)
\displaystyle
\sin^2
\frac{i}{2}
  \Bigg\} ~t.
\end{array}
\right.
\end{equation}
In the absence of torsion and when $\GG = -2$, it turns out that $(\incr
\widetilde \omega)_{\rm sec} = (\incr L_0)_{\rm sec}$, as found by
Lense-Thirring.

Using \eqref{ABCD} we rewrite \eqref{secula}.
For the nodal rate we obtain
\begin{equation}\label{Omegasecular}
(\incr \Omega)_{\rm sec} = - \frac{\GG J}{a^3(1-e^2)^{3/2}}
\Big(
1 + \mu_1
\Big) ~t,
\end{equation}
and for the longitudinal rate of the perigee
\begin{equation}\label{omegatildesecular}
   (\incr \widetilde \omega)_{\rm sec} =
\frac{2\GG J}{a^3(1-e^2)^{3/2}}
\left[
1 + \mu_2 - 3\big(1 + \mu_3\big)  \sin^2 \frac{i}{2}
\right]~t.
\end{equation}
Since $\widetilde \omega = \Omega + \omega$, for the rate of the argument
of the perigee we find
\begin{equation}\label{eq:omegasenzatilde}
(\incr \omega)_{\rm sec}=
\frac{\GG J}{a^3(1-e^2)^{3/2}}
\left[
3 + \mu_1 + 2 \mu_2 - 6 \left( 1+\mu_3\right) \sin^2 \frac{i}{2}
\right] ~t.
\end{equation}
The parameters
\begin{eqnarray*}
\mu_1 &\equiv& \frac{w_2 - w_4}{2\GG},
\\
\\
\mu_2 &\equiv&
\frac{2w_1 - w_3 + w_5}{-2\GG},
\\
\\
\mu_3 &\equiv&
\frac{4w_1 - w_2 - 2w_3 + w_4 + 2w_5}{-6\GG},
\end{eqnarray*}
measure deviations from GR. Indeed, when there is no torsion we
have $w_i=0$ for $i=1,\dots,5$. When, in addition, $\GG=-2$ the
metric is the weak field approximation of a Kerr-like metric,
and $\mu_1 = \mu_2 = \mu_3 =0$ and we get the classical
Lense-Thirring formulae \cite{LeTh:18}.
We also give the expression for the rate of the longitude at epoch,
namely
\begin{eqnarray}
(\incr L_0)_{\rm sec} = &&
- \frac{2 \GG J}{a^3(1-e^2)}
\Bigg\{
-\frac{e^2}{1 + (1-e^2)^{1/2}}
\frac{1}{(1-e^2)^{1/2}}
\Big(1+\mu_2\Big)
-\Big(1 + \mu_4\Big)
\nonumber
\\
\label{L0secular}
\\
&& + \left[
\Big(1 + \mu_1\Big)
+
\frac{3 e^2}{1 + (1-e^2)^{1/2}}\frac{1}{(1-e^2)^{1/2}}
\Big(1+\mu_3\Big)
+
2 \Big( 1 + \mu_4\Big)
\right]
\sin^2 \frac{i}{2} \Bigg\}~t,
\nonumber
\end{eqnarray}
where
\begin{equation*}
\mu_4 \equiv \frac{w_1+w_5}{-\GG}.
\end{equation*}
Note that $\mu_1,\dots,\mu_4$ do not depend on
$\torparam_1, \torparam_2, \FF, \HH$.

\section{Torsion corrections to the geodetic
precession}\label{sec:desitter}
The secular perturbations
of the orbital elements computed in
the previous sections are not the only torsion induced perturbations
that are expected.
Indeed, a further contribution due to
solar perturbation  is present, namely the geodetic precession
in presence of torsion. The corresponding perturbations of the orbital elements
have been computed in the companion paper \cite{MaBeTaDe:10} and they depend only
on the torsion parameters $t_i$.

Since we are interested in putting constraints on the
frame-dragging torsion parameters $w_1,\dots,w_5$, there is a relevant
difference between the case of GPB gyroscopes considered
in \cite{MaTeGuCa:07} and the present problem of orbits of satellites.
In \cite{MaTeGuCa:07} the average gyroscope precession rate is expressed as
\[
\left\langle\frac{d\vec S_0}{dt}\right\rangle=\vec\Omega_{\rm eff}\times\vec S_0,
\]
where $\vec S_0$ is the angular momentum of the spinning gyroscope measured by an observer
comoving with its center of mass, and the vector $\vec\Omega_{\rm eff}$
of the angular precession rate is a linear combination of $\vec\omega_O$ (the orbital angular
velocity vector of the gyroscope) and $\vec\omega_E$ (the rotational
angular velocity vector of the Earth around its axis). In $\vec\Omega_{\rm eff}$
the coefficient of $\vec\omega_O$ is a linear combination of the parameters
$t_i$,  while the coefficient of $\vec\omega_E$ is a linear combination of the parameters $w_i$.
Since the GPB satellite has a polar orbit the vectors $\vec\omega_O$ and $\vec\omega_E$ are orthogonal.
The contribution to the average precession due to $\vec\omega_O$ is the geodetic precession
of the gyroscope, while the contribution due to $\vec\omega_E$
is frame-dragging, both in the presence of torsion.
Therefore, in the GPB experiment \cite{Ev:11}, when measuring the
projections of the average precession rate of a gyroscope on the two corresponding orthogonal directions,
it turns out that the linear combinations of the $t_i$ and
of the $w_i$ torsion parameters can be constrained separately.

On the other hand, in the case of orbital motion of satellites, in the presence of torsion the
geodetic precession and the Lense-Thirring effect are superimposed as it happens in GR,
in such a way that the precessions of the orbital elements are simultaneously influenced by both effects.
In \cite{MaBeTaDe:10} it has been found that the contribution of geodetic precession
depends on a linear combination of the torsion parameters $t_i$, while the contribution of frame-dragging
computed in the previous sections depends on a linear combination of the parameters $w_i$.
It turns out that the precession of orbital elements (such as the node and the perigee)
both depend on $t_i$ and $w_i$, in such a way that
without a knowledge of the dependence of such precessions
on $t_i$, it is not possible to put constraints on the $w_i$.
The knowledge of  the dependence on $t_i$ corresponds
exactly to the knowledge of the geodetic precession
of the orbital elements in presence of torsion.

In GR it is known that the geodetic precession
is independent of the orbital elements of the satellites
(and therefore it is the same both for LAGEOS and the Moon).
This property is used in GR in order to compute an upper bound to the uncertainty in modeling
the geodetic precession, and in order to show that
the result is negligible with respect to the uncertainty in
the measurement of the Lense-Thirring effect (see \cite{CiPa:04}, Supplementary Discussion).
Such a result is important in order to extract the Lense-Thirring effect from LAGEOS data, and it is achieved
thanks to the precision of the measurement of geodetic precession by means of
lunar laser ranging (LLR) data \cite{Williams:04}.

In Section \ref{sec:simone} we will show that the
uncertainty in modeling the geodetic precession
can be neglected also in presence of spacetime torsion. In particular,
the upper bounds on the torsion parameters $t_i$ found in \cite{MaBeTaDe:10}
and recalled in the subsequent formula (\ref{stimat2}) will be useful
in order to obtain such a conclusion. This is important in order to extract the Lense-Thirring effect
from LAGEOS data also in the presence of torsion, and that will allow us to constrain
suitable linear combinations of the parameters $w_i$ separately.
Hence, in the following
we briefly need to report the results obtained in
\cite{MaBeTaDe:10}.

The geodetic  precession of orbital elements
of the satellite in the gravitational field of the Earth and the Sun
(both supposed to be nonrotating) is computed, in
a Sun-centered reference system.
It is shown that,
to the required  order of accuracy, the corresponding metric is described
by a further parameter
$\mathcal I = 2(\beta-\gamma)$, where
$\beta$ is the usual PPN parameter,
and the parametrization of the torsion tensor involves a
further parameter $t_3$ (see \cite{MaBeTaDe:10} for the details).

The secular contributions to the precessions of the node and of the perigee
due to torsion found in
\cite{MaBeTaDe:10} are the following:
\begin{equation}\label{geosec}
\begin{aligned}
(\incr \Omega^{\rm Sun})_{\rm sec} = &
\frac{1}{4}
\frac{\massasole \nu_0}{\rho}
\left(
\costante_1 - \costante_2 \frac{\nu_0}{n} \cos i
\right)
~t,
\\
\\
(\incr \widetilde \omega^{\rm Sun})_{\rm sec} = &
\frac{1}{4} \frac{
\massasole \nu_0}{\rho}
\left\{
\costante_1  + \costante_2 \frac{\nu_0}{n}
\left[
4 - \cos i - 5 \sin^2 i \sin^2 (\widetilde \omega - \Omega)
\right]
\right\} ~t,
\end{aligned}
\end{equation}
where
\begin{equation}\label{cuno}
\costante_1  \equiv 1-  \frac{\HH}{2}  + 2 \FF + 3 \torparam_2,
\qquad
\costante_2  \equiv 1 + \frac{\HH}{2} + \frac{\HH^2}{2}
- \FF - \mathcal I + \torparam_2 + 2 \torparam_3.
\end{equation}
Here $\massasole$ is the mass of the Sun,
$\nu_0$ is the revolution
angular velocity of the Earth around the Sun, and
$\rho$ is the distance of the Earth from the Sun.

Differently from the
Lense-Thirring effect, the precessions \eqref{geosec}
depend on the torsion parameters $\torparam_2$ and $\torparam_3$,
and are independent of $\torparam_4$; the parameter $\torparam_1$ is identified
using the Newtonian limit \eqref{limniuti}.

We recall that  $t_3$ and
$t_4$ enter the parametrization of torsion
at the higher order of accuracy required in the computation of
precessions \eqref{geosec}.

The perturbations \eqref{geosec} have to be superimposed to the ones
computed in  Section \ref{sec:correz}.

The first term on the right hand sides of the two
formulas in \eqref{geosec} can be interpreted as
the geodetic precession effect, when torsion is present
\cite{MaBeTaDe:10}: accordingly we set
\begin{equation}\label{geosecvere}
(\incr \Omega^{\rm geo})_{\rm sec} = (\incr \widetilde \omega^{\rm geo})_{\rm sec}  =
\frac{C_1}{4}
\frac{\massasole \nu_0}{\rho} t.
\end{equation}
In the PPN formalism we have
$$
C_1 = 2  + 4 \gamma + 3t_2, \qquad C_2 = t_2  + 2(1 -  \beta +  t_3).
$$
Using LLR data and Mercury radar ranging data respectively, the following upper bounds are given
in \cite[Section 13]{MaBeTaDe:10}:
\begin{equation}\label{stimat2}
\vert t_2\vert < 0.0128, \qquad
\vert 1-\beta + t_3\vert < 0.0286.
\end{equation}
Since for LAGEOS satellites $\frac{\nu_0}{n} \sim 4.2 \times 10^{-4}$, we have
$$
(\incr \Omega^{\rm Sun})_{\rm sec} \simeq
(\incr \Omega^{\rm geo})_{\rm sec},\qquad
(\incr  \widetilde \omega^{\rm Sun})_{\rm sec}
\simeq (\incr \widetilde \omega^{\rm geo})_{\rm sec}.
$$
Taking into account the expression of $C_1$, we have
\begin{equation}\label{soloquesta}
(\incr \Omega^{\rm geo})_{\rm sec}=
(\incr \widetilde \omega^{\rm geo})_{\rm sec} = \frac{M\nu_0}{2\rho} \left(
1+2\gamma + \frac{3}{2} t_2
\right) t.
\end{equation}
This  formula  yields the rate
of geodetic precession around an axis which is normal to the
ecliptic plane. The projection of this precession
rate on the axis of rotation of Earth is obtained by
multiplying
$(\incr \Omega^{{\rm geo}})_{\rm sec}$ by $\cos \epsilon$,
where $\epsilon=23.5$ degrees is the angle
between the Earth's equatorial plane and the ecliptic plane \cite{Huang}:
this gives the values of the geodetic precession
in a Earth-centered reference system.

\section{Constraining torsion parameters with LAGEOS}\label{sec:simone}
In this section  we describe how the LAGEOS data
can be used to extract a limit on the torsion parameters.
We will assume in the following that all metric parameters
take the same form as in the PPN formalism,
according to \eqref{postniutonianparamiters}.
Recent limits on various components of the torsion tensor, obtained
in a different torsion model based on the fact that
background torsion may violate effective local Lorentz invariance,
have been
obtained in \cite{KoRuTa}. See also \cite{HeAd}, where
constraints on possible new spin-coupled interactions using
a torsion pendulum are described.

\subsection{Constraints from nodes measurement}\label{sub:constnodes}
Here we discuss how frame dragging torsion parameters can be constrained
by the measurement of a suitable linear combination of the nodal
rates of the two LAGEOS satellites.

Equation \eqref{Omegasecular} can be rewritten as
\begin{equation}\label{nodalrate}
(\incr \Omega)_{\rm sec} = \frac{2 J}{a^3(1-e^2)^{3/2}}
\Big(-\frac{\GG}{2} - \frac{w_2 -w_4}{4} \Big) ~t =
(\incr \Omega)_{\rm sec}^{\rm GR} ~b_{\Omega},
\end{equation}
where we have defined, similarly to \cite{MaTeGuCa:07} and \cite{MaBeTaDe:10}, a
multiplicative torsion ``bias'' relative to the GR prediction as
\begin{equation}\label{Omegabias}
b_{\Omega} = \frac{(\incr \Omega)_{\rm sec}}{(\incr \Omega)_{\rm sec}^{\rm GR}}
= -\frac{\GG}{2} - \frac{w_2-w_4}{4}
= \frac{1}{2} \Big(1+\gamma + \frac{\alpha_1}{4} \Big) - \frac{w_2-w_4}{4},
\end{equation}
$(\delta \Omega)^{\rm GR}_{\rm sec} = \frac{2 J}{a^3 (1-e^2)^{3/2}}t$ being the Lense-Thirring
precession in GR.
We recall that the values of such precessions are
$31 {\rm mas/yr}$ and $31.5 {\rm mas/yr}$ for LAGEOS and LAGEOS II, respectively, where
mas/yr denotes milli-arcseconds per year.


Let us now consider the contribution of the geodetic precession
to the nodal rate.
We  write the secular contribution to the nodal rate, in a Earth-centered reference system,  in the form
\begin{equation}\label{Omegabiasdesitter}
(\incr \Omega^{{\rm geo}})_{\rm sec} \cos \epsilon =
(\incr \Omega^{{\rm geo}})_{\rm sec}^{\rm GR}  \cos \epsilon ~ \ b_{\Omega}^{\rm geo},
\end{equation}
where $b_\Omega^{\rm geo}$ depends on $\torparam_2$. Precisely, taking into account
that $(\incr \Omega^{{\rm geo}})_{\rm sec}^{{\rm GR}}
=
\frac{3 M\nu_0}{2\rho} t$
and
using \eqref{soloquesta}, we have
\begin{equation}\label{bomegageo}
b_\Omega^{\rm geo}=
\displaystyle
\frac{1}{3} (1 + 2\gamma)
+\frac{\torparam_2}{2}.
\end{equation}
Moreover, the following numerical constraints
are set on PPN parameters $\gamma$ and $\alpha_1$ by
Cassini tracking \cite{Cassini} and LLR data \cite{Wi:06},
respectively:
\begin{equation}\label{gammaalphauno}
\gamma-1=(2.1\pm2.3)\times 10^{-5}, \qquad
|\alpha_1|<10^{-4}.
\end{equation}
{}From \eqref{gammaalphauno} it follows that
the term $\frac{1}{3} (1 + 2\gamma)$ differs from 1 by a few part in $10^{-5}$.
Therefore, using \eqref{stimat2}, \eqref{bomegageo} and \eqref{gammaalphauno} we get
\begin{equation}\label{geodeticerror}
|b_\Omega^{\rm geo} - 1| \simeq |\frac{\torparam_2}{2}| < 0.0064.
\end{equation}
The measurement of the Lense-Thirring effect
in \cite{CiPa:04}, \cite{Ciufo2009} is based
on the following linear combination of the total nodal rates of the
two LAGEOS satellites:
\begin{equation}\label{totalrate}
\incr \Omega_{{\rm I}}^{\rm tot}+ \kappa  \incr \Omega_{{\rm II}}^{\rm tot},
\end{equation}
where the subscripts ${\rm I}$ and ${\rm II}$ denote LAGEOS and
LAGEOS II, respectively. Here the total nodal rate $\incr \Omega^{\rm tot}$
of a LAGEOS satellite denotes the nodal rate due to all kinds of perturbations,
both gravitational and nongravitational.
The coefficient $\kappa = 0.545$ is chosen to make the linear combination \eqref{totalrate}
independent of any contribution of the Earth's quadrupole moment $J_2$,
which describes the Earth's oblateness.

In \cite{CiPa:04} the residual (observed minus calculated) nodal rates
$\Delta(\incr \Omega_{\rm I})$, $\Delta(\incr \Omega_{\rm II})$
of the LAGEOS satellites
are obtained analyzing nearly eleven years of laser ranging data.
The residuals are then combined according to the linear combination
$\Delta(\incr \Omega_{{\rm I}})+ \kappa \Delta(\incr \Omega_{{\rm II}})$,
analogue to \eqref{totalrate}.
The Lense-Thirring effect is set equal to zero in the calculated nodal rates.
The linear combination of the residuals, after removal of the main periodic signals,
is fitted with a secular trend which corresponds to 99\%
of the theoretical Lense-Thirring prediction of GR (see \cite{CiPa:04}, \cite{Ciufo2009} for the details):
$$
(\incr \Omega_{{\rm I}})_{\rm sec}^{\rm GR}
+\kappa (\incr \Omega_{{\rm II}})_{\rm sec}^{\rm GR} = 48.2 \ \mbox{mas/yr}.
$$
The total uncertainty of the measurement is $\pm 5\%$ of the value
predicted by GR \cite{CiPa:04}, \cite{Ciufo2009}, \cite{CiPaPe:06}.
This uncertainty is a total error budget that includes all estimated systematic errors due to gravitational
and non-gravitational perturbations, and stochastic errors.
Such a result is quoted as a $1-\sigma$ level estimate in
\cite{Io:05}, \cite{Lu:07},
though an explicit indication of this fact is missing in \cite{CiPa:04}.
Eventually, the authors allow for a total $\pm 10\%$
uncertainty to include underestimated and unmodelled error sources \cite{CiPa:04}.
In the following we assume a value of $\pm 10\%$ for the uncertainty of the measurement.

Using the upper bound \eqref{geodeticerror},
the  uncertainty
  in modeling
geodetic precession  in
the presence of torsion is
\begin{equation}\label{eq:applyresults}
|b_\Omega^{\rm geo} - 1|
\Big[(\incr \Omega_{\rm I}^{{\rm geo}})_{\rm sec}^{\rm GR}+
\kappa(\incr \Omega_{\rm II}^{{\rm geo}})_{\rm sec}^{\rm GR}
\Big]
\cos \epsilon
\leq 0.0064\,\frac{27.2}{48.2}
\Big[
(\incr \Omega_{{\rm I}})_{\rm sec}^{\rm GR}
+\kappa (\incr \Omega_{{\rm II}})_{\rm sec}^{\rm GR}
\Big],
\end{equation}
where $\left[(\incr \Omega_{\rm I}^{{\rm geo}})_{\rm sec}^{\rm GR}+
\kappa(\incr \Omega_{\rm II}^{{\rm geo}})_{\rm sec}^{\rm GR}\right]\cos \epsilon=27.2$ mas/yr is the contribution
from geodetic precession predicted by GR for LAGEOS satellites.
Compared to the $\pm$10\% uncertainty in the measurement of the Lense-Thirring effect, the uncertainty
in modeling geodetic precession can be neglected
(as in \cite{CiPa:04}, \cite{Ciufo2009}) even in the presence of spacetime torsion.
This is a consequence of the torsion limits set with the Moon and Mercury in \cite{MaBeTaDe:10}.

Then we can apply the results of \cite{CiPa:04}, \cite{Ciufo2009} to our computations with torsion, and we obtain
\begin{equation*}
\Big\vert
(\incr \Omega_{{\rm I}})_{\rm sec}+ \kappa  (\incr \Omega_{{\rm II}})_{\rm sec}
-0.99\left[(\incr \Omega_{{\rm I}})_{\rm sec}^{\rm GR}
+\kappa (\incr \Omega_{{\rm II}})_{\rm sec}^{\rm GR}\right]
\Big\vert <
0.10
\left[
(\incr \Omega_{{\rm I}})_{\rm sec}^{\rm GR}
+\kappa (\incr \Omega_{{\rm II}})_{\rm sec}^{\rm GR}
\right],
\end{equation*}
where $(\incr \Omega_{{\rm I}})_{\rm sec}$ and  $(\incr \Omega_{{\rm II}})_{\rm sec}$
are given by \eqref{nodalrate}.
Since the torsion bias $b_\Omega$ does not depend on the
orbital elements of the satellite, we have
$$
\frac{(\incr \Omega_{{\rm I}})_{\rm sec}
+\kappa (\incr \Omega_{{\rm II}})_{\rm sec}}{(\incr \Omega_{{\rm I}})_{\rm sec}^{\rm GR}
+\kappa (\incr \Omega_{{\rm II}})_{\rm sec}^{\rm GR}} = b_\Omega.
$$
Hence,
using \eqref{Omegabias}, we can constrain a linear combination of the frame-dragging
torsion parameters $w_2$, $w_4$, setting the limit
$$
\vert b_{\Omega}-0.99\vert = \Big|\frac{1}{2}\Big(\gamma - 1 + \frac{\alpha_1}{4}\Big) -
\frac{w_2-w_4}{4} + 0.01\Big| < 0.10,
$$
which is shown graphically in Figure 1, together with the other
constraints on $\gamma$ and $\alpha_1$ \cite{Wi:06}.

Taking into account the numerical constraints \eqref{gammaalphauno}
 the limit on torsion
parameters from LAGEOS becomes
\begin{equation*}
\Big|- \frac{w_2-w_4}{2} + 0.02\Big| < 0.20
\end{equation*}
which implies
\begin{equation}\label{richiamo}
-0.36 < w_2-w_4 < 0.44.
\end{equation}
The constraint (\ref{richiamo}) on the torsion parameters depends on the
quantitative assessment of the uncertainty of the measurement of the Lense-Thirring effect.
However, the value 5-10\% of the uncertainty reported in \cite{CiPa:04}
has been criticized by several authors. For example Iorio argues in \cite{Io:05} that
the uncertainty might be 15-45\%. The previous computations show that the upper bound on the quantity
$$
\Big|-\frac{w_2-w_4}{4} + 0.01\Big|
$$
is given by the uncertainty of the measurement, so that one can find the constraint
on the linear combination of the torsion parameters $w_2,w_4$ corresponding to a different value
of the uncertainty. For instance, if the
value of the uncertainty of the measurement is $\pm 50\%$, the constraint on torsion parameters becomes
\[
-1.96 < w_2-w_4 < 2.04.
\]
One of the goals of the LAGEOS, LAGEOS II, LARES\footnote{LAser RElativity
Satellite, a geodynamics mission of the Italian Space Agency (ASI) to be launched.} three-satellite
experiment, together with improved Earth's gravity field models of GRACE (Gravity
Recovery And Climate Experiment) is to improve the experimental accuracy on the
orbital Lense-Thirring effect to ``a few percent'' \cite{Ciufo2009}.

We observe that, using \eqref{eq:applyresults} the uncertainty in modeling geodetic
precession in presence of torsion amounts to about $0.4\%$ of the Lense-Thirring effect,
which is still a small contribution to a total root-square-sum error of a few percent.
Note that an improved determination of the geodetic precession has been recently achieved by GPB \cite{Ev:11}
which, unlike LAGEOS, is designed to separate the frame-dragging and geodetic
precessions by measuring two different, orthogonal precessions of its gyroscopes.

\begin{figure}[!ht]
\begin{center}
\label{LimitPlot}
\includegraphics[width=0.8\textwidth]{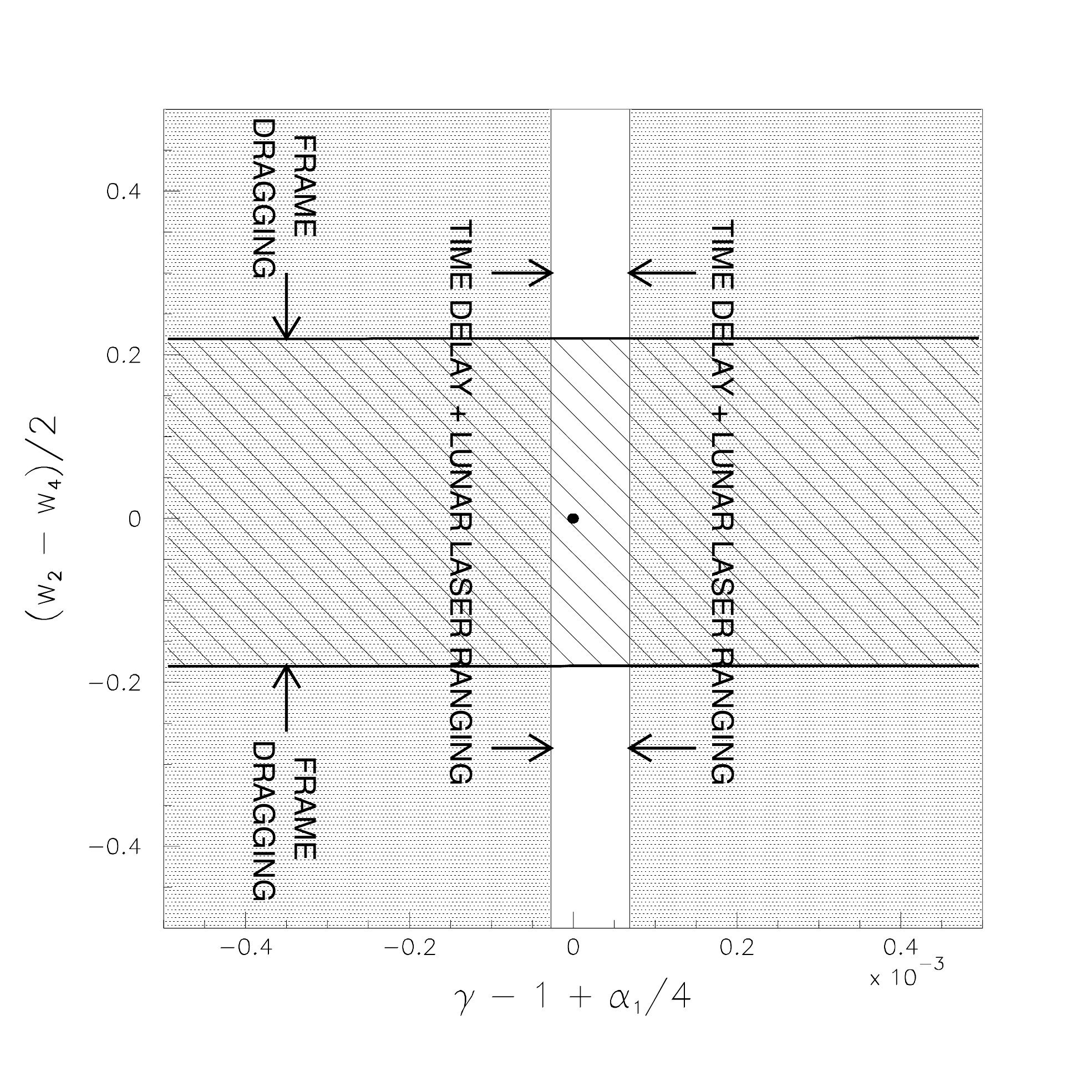}
\caption{constraints on PPN parameters ($\gamma$, $\alpha_1$) and on frame-dragging
torsion parameters ($w_2, w_4$) from solar system tests. The grey area is the region
excluded by lunar laser ranging and Cassini tracking.
The LAGEOS nodes measurement of the Lense-Thirring effect \cite{CiPa:04}, \cite{Ciufo2009} excludes values
of $(w_2 - w_4)/2$ outside the hatched region. General Relativity corresponds to $\gamma=1$, $\alpha_1=0$
and all torsion parameters = 0 (black dot).}
\end{center}
\end{figure}

In the case of GPB, the torsion bias for the precession of a gyroscope is
 \cite{MaTeGuCa:07}
 \begin{eqnarray*}
 -\frac{\GG}{2} - \frac{w_1+w_2-w_3-2w_4+w_5}{2}.
\end{eqnarray*}
  This formula (the analogue
of the right hand side of  equation \eqref{Omegabias})
involves a linear combination of all frame-dragging torsion parameters. Such a linear
combination  can be constrained from GPB data.
Since LAGEOS and GPB are sensitive to different linear combinations,
together they can put more stringent torsion limits.

After taking into account  the contribution of the geodetic precession,
the combined constraints from gyroscope and orbital Lense-Thirring experiments are
effective probes to search for the experimental signatures of spacetime torsion.
In this sense, LAGEOS and GPB are to be considered complementary
frame-dragging and, at the same time, torsion experiments, with the
notable difference that GPB measures also the geodetic precession.

\subsection{Constraints from nodes and perigee measurement}\label{sub:constnoperig}
In this section we discuss how frame dragging torsion parameters can be constrained
by the measurement of a linear combination of the nodal
rates of  LAGEOS and LAGEOS II and the perigee rate
of LAGEOS II.

Similarly to the previous section, we define a multiplicative torsion
``bias'' relative to the GR prediction also for the
rate of the argument of the perigee \eqref{eq:omegasenzatilde}:
\begin{equation*}
b_{\omega} =
\frac{(\incr \omega)_{{\rm sec}}}{(\incr\omega)^{{\rm GR}}_{\rm sec}}
= - \frac{\GG}{6 \cos i}~  \left[3 + \mu_1 + 2\mu_2 - 6(1+ \mu_3) \sin^2 \frac{i}{2}
\right],
\end{equation*}
$(\incr\omega)^{{\rm GR}}_{\rm sec} = -\frac{6J \cos i}{a^3(1-e^2)^{3/2}}t$ being the
Lense-Thirring precession in GR: we recall that the value of this precession is
$-57 {\rm mas}/{\rm yr}$ for LAGEOS II. In the following, the torsion bias $b_\omega$
is referred to LAGEOS II.

Using the values of $\mu_1$, $\mu_2$ and $\mu_3$ given in Section \ref{sec:correz}
we find
\begin{equation}\label{bomega}
b_\omega = -\frac{\GG}{2} + \frac{4w_1 - w_2 - 2 w_3 + w_4 + 2 w_5}{12}.
\end{equation}
The measurement of the Lense-Thirring effect
in \cite{Ciufo98} is based
on the following linear combination of the residuals of the nodes
of
LAGEOS and LAGEOS II and of the perigee  of LAGEOS II:
\begin{equation}\label{totalrateperig}
\Delta(\incr \Omega_{{\rm I}})+ c_1 \Delta(\incr \Omega_{{\rm II}}) + c_2
\Delta(\incr
 \omega_{\rm II}),
\end{equation}
where
the coefficients $c_1= 0.295$
and $c_2 =-0.35 $ are chosen to make the linear combination \eqref{totalrateperig}
independent of the first two even zonal harmonic coefficients $J_2$ and $J_4$,
and of their uncertainties.

In \cite{Ciufo98} the residuals
are obtained analyzing four years of laser ranging data, and then
 combined according to the linear combination  \eqref{totalrateperig}.
The Lense-Thirring effect is set equal to zero in the calculated rates of the nodes and of the perigee.
The linear combination of the residuals, after removal of the main periodic signals and of small
observed inclination residuals,
is fitted with a secular trend which corresponds to $1.1$ times
the theoretical Lense-Thirring prediction of GR (see \cite{Ciufo98} for the details):
$$
(\incr \Omega_{{\rm I}})_{\rm sec}^{\rm GR}
+c_1 (\incr \Omega_{{\rm II}})_{\rm sec}^{\rm GR}  + c_2
(\incr  \omega_{{\rm II}})_{\rm sec}^{\rm GR} = 60.2 \ \mbox{mas/yr}.
$$
The total uncertainty of the measurement found in \cite{Ciufo98} is
 $\pm 20\%$ of the value predicted by GR.
This uncertainty is a total error budget that includes all the estimated systematic errors due to gravitational
and non-gravitational perturbations.
Such a result is quoted as a $1-\sigma$ level estimate in \cite{Io:05},
though an explicit indication of this fact is missing in \cite{Ciufo98}.

The contribution to the
uncertainty of the measurement due to nongravitational perturbations, mainly
thermal perturbative effects, on the perigee of LAGEOS II,
 amounts to $13\%$ of the value predicted by GR.
 In \cite{Lu:02} such an estimate is confirmed,
however the author, when considering more pessimistic assumptions on
some thermal effects, estimates that the contribution of nongravitational
perturbations to the total uncertainty does not exceed the  $28\%$
of the GR value. Here we will follow this more conservative estimate.
Inserting this value in the estimate of the total
uncertainty computed in \cite{Ciufo98} yields a
total  root-square-sum error of $32\%$ of the GR value.

For reasons similar to the ones discussed in the previous
section, we are allowed to neglect the uncertainty in modeling
 the geodetic precession in presence of torsion.
Then we can apply the results of \cite{Ciufo98} to our computations with torsion, and
we obtain
\begin{equation*}
\begin{aligned}
& \Big\vert(\incr \Omega_{{\rm I}})_{\rm sec}+ c_1  (\incr \Omega_{{\rm II}})_{\rm sec}
+ c_2  (\incr \omega_{{\rm II}})_{\rm sec}
-1.1 \left[(\incr \Omega_{{\rm I}})_{\rm sec}^{\rm GR}
+c_1 (\incr \Omega_{{\rm II}})_{\rm sec}^{\rm GR}
+ c_2  (\incr \omega_{{\rm II}})_{\rm sec}^{\rm GR}
\right]\Big\vert
\\
\\
& \qquad \qquad \qquad \qquad \qquad \qquad <
0.32
\left[
(\incr \Omega_{{\rm I}})_{\rm sec}^{\rm GR}
+c_1(\incr \Omega_{{\rm II}})_{\rm sec}^{\rm GR}
+ c_2  (\incr \omega_{{\rm II}})_{\rm sec}^{\rm GR}
\right].
\end{aligned}
\end{equation*}
A direct computation gives
\begin{equation}\label{sstima}
\left\vert
(1-K) b_\Omega + K b_\omega
-1.1 \right\vert < 0.32,
\end{equation}
where
$$
\begin{aligned}
K=  &
\frac{
c_2(\incr \omega_{{\rm II}})_{\rm sec}^{\rm GR}}{(\incr \Omega_{{\rm I}})_{\rm sec}^{\rm GR}
+c_1(\incr \Omega_{{\rm II}})_{\rm sec}^{\rm GR}
+ c_2  (\incr \omega_{{\rm II}})_{\rm sec}^{\rm GR}}= 0.33.
\end{aligned}
$$
Inserting in \eqref{sstima} the expressions of $b_\Omega$ and $b_\omega$ given in \eqref{Omegabias}, \eqref{bomega}
and taking into account that  $\GG\simeq -2$ by formula \eqref{gammaalphauno}, we obtain
$$
-0.22 <
- \frac{w_2-w_4}{4} +
K
\left(\frac{2w_1 +w_2 -  w_3 - w_4 +  w_5}{6}\right)
 < 0.42.
$$
Using the value of $K$ we finally deduce
\begin{equation}\label{unsenepolepiu}
-0.22 <
0.11 w_1 - 0.20 w_2 - 0.06 w_3
+ 0.20 w_4+0.06 w_5
 < 0.42,
\end{equation}
%
which is shown graphically in Figure 2, together with the other
constraints on $\gamma$ and $\alpha_1$ \cite{Wi:06}.

\begin{figure}[!ht]
\begin{center}
\label{LimitPlot2}
\includegraphics[width=0.8\textwidth]{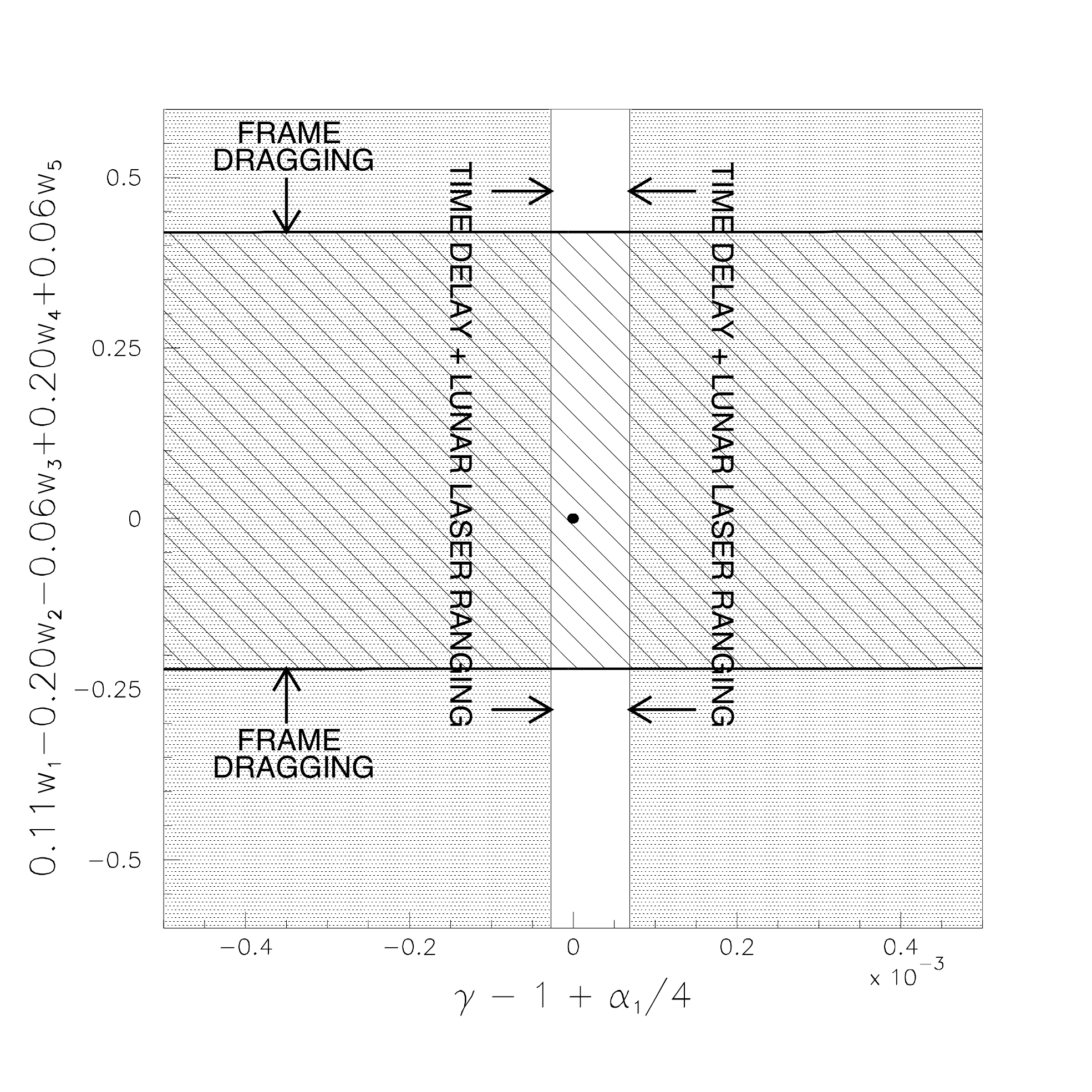}
\caption{constraints on PPN parameters ($\gamma$, $\alpha_1$) and on frame-dragging
torsion parameters ($w_1, w_2, w_3, w_4, w_5$) from solar system tests. The grey area is the region
excluded by lunar laser ranging and Cassini tracking.
The LAGEOS nodes and perigee measurement of the Lense-Thirring effect \cite{Ciufo98}, \cite{Lu:02}
excludes values of $0.11 w_1 - 0.20 w_2 - 0.06 w_3  + 0.20 w_4+0.06 w_5$
outside the hatched region. General Relativity corresponds to $\gamma=1$, $\alpha_1=0$
and all torsion parameters = 0 (black dot).}
\end{center}
\end{figure}

The constraint \eqref{unsenepolepiu} on the
linear combination of the frame-dragging parameters
is rather weak, due to the uncertainty on
the nongravitational perturbations. Notice  that
the coefficients in front of $w_3$ and $w_5$ are
of an order of magnitude smaller than
the coefficients of the other parameters, so that the constraint
on $w_3$ and $w_5$ is even looser.

Thermal thrusts (TTs) are the main source of non-gravitational perturbations \cite{Lu:02}.
One of the main drivers of LAGEOS TTs is the thermal relaxation time $\tau_{\rm CCR}$ of
its fused silica cube corner retroreflectors \cite{INFNPaper:06}, which has been characterized
in
laboratory-simulated space conditions at the INFN-LNF ~Satellite/lunar  laser ranging
Characterization Facility (SCF) \cite{ILRS08}, \cite{INFNPaper:10}, \cite{MAGIA}.
The measurements
of LAGEOS  $\tau_{\rm CCR}$  in a variety of thermal conditions provide the basis
for possibly reducing the uncertainty on the thermal perturbative effects.
As a consequence, the constraint \eqref{unsenepolepiu} could be improved.

The constraint (\ref{unsenepolepiu}) on the torsion parameters depends on the
quantitative assessment of the uncertainty of the measurement of the Lense-Thirring effect.
Again, the value $\pm 20\%$ of the uncertainty reported in \cite{Ciufo98}
has been criticized by various authors. For example Ries, Eanes and Tapley
argue in \cite{RiEaTa:03} that the uncertainty is at best in the 50-100\% range.
The uncertainty of the measurement yields the upper bound on the right-hand side
of the estimate (\ref{sstima}). Hence, one can find the constraint
on the linear combination of the torsion parameters $w_i$ corresponding to a different value
of the uncertainty as it has been discussed in Section \ref{sub:constnodes}.

We recall that in
\cite{MaTeGuCa:07} an upper bound on the combination $\vert w_1+w_2-w_3-2w_4+w_5\vert$
is given. This constrains  the torsion
parameters within two parallel hyperplanes in a five-dimensional space.
If we couple this bound with our two
estimates \eqref{richiamo} and \eqref{unsenepolepiu},
we obtain that
 $w_1,\dots, w_5$
are constrained to lye in
a five-dimensional set, which is unbounded only
along two directions. Hence, coupling GPB with SLR measurements
significantly reduces the degrees of freedom on the frame-dragging parameters.

We conclude this section by observing that the recently approved JUNO mission to Jupiter \cite{Matousek}
will make it possible, in principle, to attempt a measurement of the Lense-Thirring effect
through the JUNO's node, which would be displaced by about 570 metres over the mission
duration of one year \cite{Iorio-Juno}. Hence, such a
mission yields an opportunity for a possible improvement of the costraints on torsion parameters.

\section{Conclusions}\label{sec:conclus}
We have applied the framework recently developed in
\cite{MaTeGuCa:07} for GR with torsion, to the computation of
the slow orbital motion of a satellite in the field generated by the
Earth. Starting from the autoparallel trajectories, we computed the
corrections to the classical orbital Lense-Thirring effect in the
presence of torsion. By using perturbation theory, we have found
the explicit dependence of the secular variations of the longitudes
of the node and of the perigee on the frame-dragging torsion parameters.
The  LAGEOS
nodes measurements \cite{CiPa:04}, \cite{Ciufo2009} and the LAGEOS nodes and perigee
measurements \cite{Ciufo98}, \cite{Lu:02} of the Lense-Thirring effect
can be used to place constraints on torsion parameters, which are different
and complementary to those set by GPB.

\section{Appendix}\label{sec:append}
Under spherical axisymmetry assumptions, the metric tensor
$g_{\secondoindice\terzoindice}$ can be parametrized to first order
as follows \cite{MaTeGuCa:07}:
\begin{equation}\label{dsFHG}
ds^2 = -
\left[1+ \HH \frac{m}{r}\right] dt^2 +
\left[1+ \FF \frac{m}{r}\right] dr^2 +
r^2  (d \theta^2 + \sin^2\theta~ d\phi^2) + 2 \GG \frac{J}{r} \sin^2\theta ~dt d\phi,
\end{equation}
where
$\HH, \mathcal
F, \mathcal
G$ are
three dimensionless parameters that
can be
immediately related to the Parametrized Post Newtonian (PPN) parameters:
\begin{equation}
\label{postniutonianparamiters}
\HH = -2, \qquad
\FF = 2 \gamma, \qquad
\GG =
- \left(1 + \gamma + \frac{\alpha_1}{4}\right).
\end{equation}
Here we follow the notation of the paper \cite{MaTeGuCa:07}, instead
of the PPN notation.
This will be useful in Section \ref{sec:desitter}.

The nonvanishing components of the torsion tensor
are:
\begin{eqnarray}\label{Sfirst}
S_{tr}^{\ \ t}&=& \torparam_1\frac{m}{2r^2},
\nonumber
\\
S_{r\theta}^{\ \ \theta}&=&S_{r\phi}^{\ \ \phi}=\torparam_2\frac{m}{2r^2},
\nonumber
\\
S_{r\phi}^{\ \ \ \! t}&=&w_1\frac{J}{2r^2} \sin^2\theta,
\nonumber
\\
S_{\theta\phi}^{\ \ \ \! t}&=&w_2\frac{J}{2r}\sin \theta \cos \theta,
\nonumber
\\
S_{t\phi}^{\ \ r}&=&w_3\frac{J}{2r^2} \sin^2\theta,
\\
S_{t\phi}^{\ \ \theta}&=&w_4\frac{J}{2r^3}\sin \theta \cos \theta,
\nonumber
\\
S_{t r}^{\ \ \phi}&=&w_5\frac{J}{2r^4},
\nonumber
\\
S_{t \theta}^{\ \
\phi}&=&-w_4\frac{J}{2r^3}\frac{\cos\theta}{\sin \theta }.
\nonumber
\end{eqnarray}
The expression of
the nonvanishing components of
the connection approximated to first order in
$\em=m/r$, $\ea=J/(mr)$ and $\em\ea=J/r^2$ is
the following \cite{MaTeGuCa:07}:
\begin{eqnarray*}
\Gamma^{t}_{\ tr}&=&\frac{1}{2r}\left(2 \torparam_1 -\HH\right)\em,
\\
\Gamma^{t}_{\ rt}&=&-\frac{\HH}{2r}\,\em,
\\
\Gamma^{t}_{\ r\phi}&=&\frac{1}{2}(3\GG+(w_1-w_3-w_5))\sin^2 \theta\,\em\ea,
\\
\Gamma^{t}_{\ \phi r}&=&\frac{1}{2}(3\GG-(w_1+w_3+w_5))\sin^2 \theta\,\em\ea,
\\
\Gamma^{t}_{\ \theta \phi}&=&\frac{1}{2} w_2 r
\sin \theta \cos \theta \,\em\ea,
\\
\Gamma^{r}_{\ tt}&=&\frac{1}{2r}\left(2 \torparam_1 -\HH\right)\em,
\\
\Gamma^{r}_{\ rr}&=&-\frac{\FF}{2r}\,\em,
\\
\Gamma^{r}_{\ \theta\theta}&=& -r + (\torparam_2+\FF)r\,\em,
\\
\Gamma^{r}_{\ \phi\phi}&=&-r\sin^2 \theta+\frac{1}{r}(\FF +
\torparam_2)\sin^2 \theta\,\em,
\\
\Gamma^{r}_{\ t\phi}&=&\frac{1}{2}(\GG-(w_1-w_3+w_5))\sin^2 \theta\,\em\ea,
\\
\Gamma^{r}_{\ \phi t}&=&\frac{1}{2}(\GG-(w_1+w_3+w_5))\sin^2 \theta\,\em\ea,
\end{eqnarray*}
\begin{eqnarray*}
\Gamma^{\theta}_{\ t\phi}&=&-\frac{1}{2r}(2\GG+(w_2-2w_4))
\sin \theta \cos \theta \,\em\ea,
\\
\Gamma^{\theta}_{\ \phi t}&=&-\frac{1}{2r}(2\GG+ w_2)
\sin \theta \cos \theta \,\em\ea,
\\
\Gamma^{\theta}_{\ r\theta}&=&\Gamma^{\phi}_{\ r\phi}=\frac{1}{r},
\\
\Gamma^{\theta}_{\ \theta r}&=&\Gamma^{\phi}_{\ \phi r}=\frac{1}{r}-\frac{1}{r} \torparam_2\, \em,
\\
\Gamma^{\theta}_{\ \phi\phi}&=&-\sin \theta
\cos \theta,
\\
\Gamma^{\phi}_{\ t r}&=&-\frac{1}{2r^2}(\GG-(w_1-w_3+w_5))\,\em\ea,
\\
\Gamma^{\phi}_{\ r t}&=&-\frac{1}{2r^2}(\GG-(w_1-w_3-w_5))\,\em\ea,
\\
\Gamma^{\phi}_{\ t \theta}&=&\frac{1}{2r}(2\GG+(w_2-2w_4))
\frac{\cos \theta}{\sin \theta}\,\em\ea,
\\
\Gamma^{\phi}_{\ \theta t}&=&\frac{1}{2r}(2\GG+ w_2)
\frac{\cos \theta}{\sin \theta}\,\em\ea,
\\
\Gamma^{\phi}_{\ \theta\phi}&
=&\Gamma^{\phi}_{\ \phi\theta}=\frac{\cos\theta}{\sin\theta}.
\end{eqnarray*}

\section*{Acknowledgments}

We thank the University of Roma ``Tor Vergata'', CNR and INFN for
supporting this work. We thank I. Ciufolini for suggesting this
analysis after the publication of the paper by MTGC \cite{MaTeGuCa:07}, and
B. Bertotti and A. Riotto for  useful advices.


\end{document}